\documentstyle[epsfig]{mn}

\newif\ifAMStwofonts
\AMStwofontstrue

\newcommand{\be}{\begin{equation}}
\newcommand{\ee}{\end{equation}}

\ifoldfss
  \ifCUPmtlplainloaded \else
    \NewTextAlphabet{textbfit} {cmbxti10} {}
    \NewTextAlphabet{textbfss} {cmssbx10} {}
    \NewMathAlphabet{mathbfit} {cmbxti10} {} 
    \NewMathAlphabet{mathbfss} {cmssbx10} {} 
  \fi
  \ifAMStwofonts
    \ifCUPmtlplainloaded \else
      \NewSymbolFont{upmath} {eurm10}
      \NewSymbolFont{AMSa} {msam10}
      \NewMathSymbol{\upi}     {0}{upmath}{19}
      \NewMathSymbol{\umu}     {0}{upmath}{16}
      \NewMathSymbol{\upartial}{0}{upmath}{40}
      \NewMathSymbol{\leqslant}{3}{AMSa}{36}
      \NewMathSymbol{\geqslant}{3}{AMSa}{3E}

      \let\leq=\leqslant 
       
    \fi
  \fi
\fi 

\ifnfssone
  \newmathalphabet{\mathit}
  \addtoversion{normal}{\mathit}{cmr}{m}{it}
  \addtoversion{bold}{\mathit}{cmr}{bx}{it}
  \newmathalphabet{\mathbfit} 
  \addtoversion{normal}{\mathbfit}{cmr}{bx}{it}
  \addtoversion{bold}{\mathbfit}{cmr}{bx}{it}
  \newmathalphabet{\mathbfss} 
  \addtoversion{normal}{\mathbfss}{cmss}{bx}{n}
  \addtoversion{bold}{\mathbfss}{cmss}{bx}{n}
  \ifAMStwofonts
    \ifCUPmtlplainloaded \else
      \UseAMStwoboldmath
      \makeatletter
      \new@mathgroup\upmath@group
      \define@mathgroup\mv@normal\upmath@group{eur}{m}{n}
      \define@mathgroup\mv@bold\upmath@group{eur}{b}{n}
      \edef\UPM{\hexnumber\upmath@group}
      \new@mathgroup\amsa@group
      \define@mathgroup\mv@normal\amsa@group{msa}{m}{n}
      \define@mathgroup\mv@bold\amsa@group{msa}{m}{n}
      \edef\AMSa{\hexnumber\amsa@group}
      \makeatother
      \mathchardef\upi="0\UPM19
      \mathchardef\umu="0\UPM16
      \mathchardef\upartial="0\UPM40
      \mathchardef\leqslant="3\AMSa36
      \mathchardef\geqslant="3\AMSa3E

      \let\leq=\leqslant 

    \fi
  \fi
\fi 

\ifnfsstwo
  \DeclareMathAlphabet{\mathbfit}{OT1}{cmr}{bx}{it}
  \SetMathAlphabet\mathbfit{bold}{OT1}{cmr}{bx}{it}
  \DeclareMathAlphabet{\mathbfss}{OT1}{cmss}{bx}{n}
  \SetMathAlphabet\mathbfss{bold}{OT1}{cmss}{bx}{n}
  \ifAMStwofonts
    \ifCUPmtlplainloaded \else
      \DeclareSymbolFont{UPM}{U}{eur}{m}{n}
      \SetSymbolFont{UPM}{bold}{U}{eur}{b}{n}
      \DeclareSymbolFont{AMSa}{U}{msa}{m}{n}
      \DeclareMathSymbol{\upi}{0}{UPM}{"19}
      \DeclareMathSymbol{\umu}{0}{UPM}{"16}
      \DeclareMathSymbol{\upartial}{0}{UPM}{"40}
      \DeclareMathSymbol{\leqslant}{3}{AMSa}{"36}
      \DeclareMathSymbol{\geqslant}{3}{AMSa}{"3E}

      \let\leq=\leqslant 

    \fi
  \fi
\fi 

\ifCUPmtlplainloaded \else
  \ifAMStwofonts \else 
    \def\upi{\pi}
    \def\umu{\mu}
    \def\upartial{\partial}
  \fi
\fi

\begin{document}

\title[Minkowski Functionals]{\bf Topology of the Las Campanas Redshift Survey}
 
\author[H. Trac et al.]{Hy Trac$^{1,2}$, Dimitris Mitsouras$^3$, Paul Hickson$^1$, and Robert Brandenberger$^{4,1}$\\
$^1$ Department of Physics and Astronomy,
        University of British Columbia, Vancouver, B.C.  V6T 1Z1, Canada \\
   	$^2$ Astronomy Department, University of Toronto, Toronto, ON M5S          3H8, Canada\\ 
$^3$ Lab For Computer Science, Massachusetts Institute of Technology, Cambridge, MA 02139, USA\\
$4$ Physics Department, Brown University, Providence, RI 02912, USA} 

\maketitle

\begin{abstract}  
The topology of the Las Campanas Redshift Survey is analyzed using Minkowski functional statistics, taking into account the selection effects of the survey. The results are compared with the predictions of some toy models of structure formation, including the Standard Cold Dark Matter Model and topological defect-based models. All of the toy models have a scale invariant primordial spectrum of perturbations, but quite different topologies. The statistics can discriminate between the predictions of the models with high significance.  Amongst the four Minkowski functionals, the integrated mean curvature statistic appears to be the most powerful discriminant, followed by the genus statistic.  None of the models considered gives an acceptable fit to the data.

\end{abstract}

\begin{keywords}
cosmology: theory -- large scale structure, redshift surveys
\end{keywords}

\noindent BROWN-HET- \hfill January 2000
 
\section{Introduction} 

At the present time the topology of the large-scale structure (LSS) of the Universe is not well known. Some surveys show pronounced coherent planar structures such as the ``Great Wall" (see e.g. de Lapparent et al. 1986), others show evidence of filaments (see e.g. Haynes and Giovanelli 1986), and yet others appear dominated by superclusters and voids (see e.g. Kirshner et al. 1981). In many cases, however, these conclusions are reached by visual inspection rather than by a rigorous statistical analysis.

The topology of the LSS is very important for cosmology. Simple inflation-based models predict a Gaussian distribution of the density perturbations (see e.g. Mukhanov et al. 1992, for a comprehensive review), whereas topological defect-based models predicted a non-Gaussian density field (see e.g. Brandenberger 1994, for a review). These models to a first approximation all predict a scale-invariant spectrum of primordial perturbations; and therefore the predictions of the models cannot be differentiated by means of power spectra and two-point functions. However, defect models can be differentiated amongst one-another as well as from inflation-based models in terms of the topology of LSS which they yield. 

The currently favored simple inflation-based theories predict a Gaussian distribution of the density fluctuations. Nonlinear evolution and biasing are expected to produce secondary non-Gaussianities, but of a type very different from the primary non-Gaussian signals predicted in defect models. For example, the genus curve (one of the Minkowski functionals used in our work) of a simulated SLOAN redshift survey is indistinguishable from that of a Gaussian distribution (Colley et al. 1999). 

The nontrivial topological signatures of most defect models are impossible to detect in the current CMB experiments (and hard to detect even in future experiments such as MAP) since due to the central limit theorem the non-Gaussian effects disappear at the large angular scales accessible (see e.g. Moessner et al. 1994, for a specific study). Conversely, the topology of the LSS may be less sensitive to complicated biasing effects than the topology on smaller scales, thus rendering the study of the topology of the LSS an interesting endeavor.

Recently, the use of the {\it Minkowski functional statistics} has been suggested as a useful way to study the topology of structure in the Universe (Mecke et al. 1994, Schmalzing and Buchert 1997). It has been shown (Mitsouras 1998, Mitsouras et al. 1999) that Minkowski functionals can differentiate between models with identical power spectra such as defect toy models and inflation-based models. Amongst the four Minkowski functionals $M0 - M3$, the $M2$ statistic (integrated mean curvature) was shown to be a better discriminator than the frequently used genus statistic $M3$. Minkowski functionals have been applied to Abell galaxy clusters surveys (Kerscher et al. 1997) and to the IRAS catalog (Kerscher et al. 1998), among others, and to numerical simulations (Schmalzing et al. 1999). 

Another set of useful statistics to explore the morphology of the large-scale structure are the {\it structure functions} (Babul and Starkman 1992, Sathyaprakash  et al. 1996) which have e.g. been applied to the 1.2 Jy IRAS catalog (Sathyaprakash et al. 1998). These statistics are based on ratios of eigenvalues of the inertia tensor of the mass distribution and carry information different from that of the Minkowski functionals. On the other hand, the {\it Shapefinder statistics} (Sahni et al. 1998) are ratios of Minkowski functionals. They have very recently (Bharadwaj et al. 2000) been used to analyze the LCRS data set (without taking into account the important selection effects), finding evidence for filamentarity.   

The deepest wide field redshift survey to date is the Las Campanas Redshift Survey (LCRS) (Shectman et al 1996). An initial analysis (Doroshkevich et al 1996) indicated that the LCRS field was dominated by planar structures, with higher density filaments at their intersections. In contrast, the analysis of Mitsouras (Mitsouras 1998) using the Minkowski functional statistics detected no deviations from a Gaussian distribution. Neither analysis, however, took into account the complicated selection effects intrinsic to the LCRS survey.  

In this paper, we calculate the Minkowski functionals for the LCRS, taking careful account of the nontrivial selection effects of that survey (except for the limited surface brightness selection effect). We compare the resulting statistics with the predictions of some toy models of structure formation: the 
Standard Cold Dark Matter (SCDM) model as a representative of the class of inflation-based models, and three toy defect models (based on string wakes, string filaments, and textures, respectively). Our aim in this paper is to focus on the potential of the LCRS survey to differentiate between different models of structure formation with identical power spectra rather than on studying specific COBE normalized models. We find that no model comes close to the data.
In work in progress we are studying a class of COBE-normalized inflation-based models with the goal of discovering whether some members of this class come closer to explaining the observed topology than the SCDM model.

\section{Data Analysis}

The Las Campanas Redshift Survey (LCRS) is a fiber-optic survey consisting of 26418 redshifts of galaxies.  $1.5^{\circ} \times 1.5^{\circ}$ spectroscopic fields were observed to pave 700 deg$^{2}$ of sky in six strips, each roughly $1.5^{\circ} \times 80^{\circ}$.  Three strips centered on $\delta = -3^{\circ}$, $-6^{\circ}$, and $-12^{\circ}$ are in the North galactic cap (NGC) and three centered on $\delta = -39^{\circ}$, $-42^{\circ}$, and $-45^{\circ}$ are in the South galactic cap (SGC).  The LCRS probed to redshifts $z \simeq 0.2$ with mean redshift ${\bar z} \simeq 0.1$ or equivalently to distances of 600 $h^{-1}$Mpc.  Approximately 70\% of galaxies having magnitudes in the range $15.0 \leq m_{r} < 17.7$ were measured spectroscopically.  For our analysis, we only consider galaxies out to a redshift $z<0.15$ or distances of 450 $h^{-1}$Mpc or less.

The selection criteria for the LCRS are more complex than a simple magnitude limit and have their origin in the technique of the fiber-optic measurements.  Galaxies were selected from a CCD-based catalog based on four selection criteria.  For the first 20\% of survey galaxies, redshifts were measured using a 50-fiber spectrograph system and the final 80\% using an improved 112-fiber system.  Galaxies were first selected by applying both faint and bright isophotal magnitude limits, with nominal values
\begin{equation}
\begin{array}{l}
m_1\leq m<m_2: \\
\quad\left\{\begin{array}{l}
m_1=16.0,\ m_2=17.3, \ {\rm 50-fiber \ data} \\
m_1=15.0,\ m_2=17.7, \ {\rm 112-fiber \ data}
\end{array} \right. \ .
\end{array}
\end{equation}
The second criterion selects against low surface brightness (LSB) galaxies by imposing the nominal central surface brightness cutoff
\begin{equation}
\begin{array}{l}
m_c<m_{cen}-0.5(m_2-m): \\
\quad\left\{\begin{array}{l}
m_{cen}=18.15,\ {\rm 50-fiber \ data} \\
m_{cen}=18.85,\ {\rm 112-fiber \ data}
\end{array} \right. \ ,
\end{array}
\end{equation}
where the central magnitude $m_c$ corresponds to the flux within a 2 pixel radius of the centroid of the galaxy.  There are small field-to-field variations from the nominal photometric limits which were made in response to variations in sky brightness and in seeing.  The third criterion restricts the maximum number of galaxies in a given field to the number of fibers available during observation of that particular field.  Lastly, the fourth criterion imposes a minimum separation of 55$^{\prime\prime}$ between galaxies because mechanical constraints prevent the fibers from being placed any closer together.

\section{Models}

There are two main classes of theories for structure formation which may possibly be able to account for the origin of primordial perturbations and observed large-scale structure in the universe.  The first are inflation-based models in which quantum fluctuations are generated at an early stage of inflation on microphysical scales, and the wavelength of these perturbations subsequently grows exponentially and thus gives rise to density perturbations on cosmological scales (Mukhanov and Chibisov 1981, see also Press 1980, Lukash 1980, Chibisov \& Mukhanov 1980, Sato 1981, Chibisov \& Mukhanov 1982 for related work).  Simple inflation-based models predict a scale-invariant spectrum of adiabatic perturbations with random phases, each having a Gaussian distribution.  The second class is based on topological defects, formed during a symmetry breaking phase transition in the early universe, giving rise to seeds for structure formation (Kibble 1976, Zel'dovich 1980, Vilenkin 1981).  Topological defects represent regions in space with trapped energy density and they result in a roughly scale-invariant spectrum of non-adiabatic perturbations with non-Gaussian phases.  In this paper, we compare the inflation-based ``standard" cold dark matter (SCDM) model with topological defect models like cosmic strings and textures. \footnote{After this work was completed, the results of the BOOMERANG (de Bernardis et al 2000) and MAXIMA-I (Hanany et al. 2000) CMB experiments were announced, which with high statistical significance demonstrate the existence of a well-defined and narrow first acoustic peak in the angular power spectrum, thus ruling out topological defects as the main source of density fluctuations (see e.g. Albrecht et al 1999 and references therein). It is, however, important to keep in mind that defects may still contribute a significant amount to the primordial power spectrum (see e.g. Bouchet et al. 2000), and that the resulting non-Gaussian signals in the large-scale structure of the Universe may well be important.}

\subsection{Standard Cold Dark Matter Model (SCDM)}

In the SCDM model, the dominant constituent of the universe is cold dark matter and the standard model is described by a density parameter $\Omega=1$, Hubble constant $H_\circ=100h$ km/s/Mpc with $h=0.5$, and cosmological constant $\Lambda=0$.  This specific model conflicts with observations if we compare the normalization of the power spectrum of density perturbations with that of the CMB fluctuations.  Other inflationary models like tilted CDM, $\Lambda$CDM, and mixed dark matter models have the potential to account for observations.  Since we are primarily concerned with the power of the Minkowski functionals to differentiate between the signatures of primordial Gaussian and non-Gaussian phases, we have chosen to work with the standard model.  We have also modified the standard value of the Hubble constant from $h = 0.5$ to $h = 0.7$ to reflect some of the more recent measurements.

In the SCDM simulation, 192$^3$ particles are evolved in a cube of comoving length of 630 $h^{-1}$Mpc using the Bertschinger \& Gelb (1991) $P^3M$ N-body code.  The normalization of the primordial power spectrum was chosen such that $\sigma_8=0.95$ to agree with the observed clustering (Pen 1998). A COBE normalized model would require $\sigma_8=1.22$.

\subsection{Cosmic Strings}

Cosmic strings are linear topological defects which arise if the vacuum manifold $\cal M$ is not simply connected.  At the time of the phase transition, a network of strings is formed.  The phase transition takes place on a short time scale $\Delta t<H^{-1}$, and will lead to correlation regions of radius $\xi<t$.  The strings are either infinite in length or closed loops.  Infinite strings give rise to filaments or wakes while loops form seeds for spherical accretion of matter and radiation.  In our study, we are concerned only with the infinite strings, but realistic cosmic string models should contain a mixture of both.

In the string wake model, strings are straight on the scale of the Hubble radius and the string tension $|p|$ equals the mass per unit length $\mu$.  As a result, the strings will move relativistically and exert no local gravitational force.  The strings will have a large transverse velocity $v$ and give rise to a planar velocity perturbation.  A string at time $t_{i}$ produces a wake of comoving length $v\xi(t_i)z(t_i)$ and width $\xi(t_i)z(t_i)$.  The wake thickness can be calculated using the Zel'dovich approximation (Zel'dovich 1970) and will depend on the nature of the dark matter since free streaming has an effect.  For hot dark matter, the thickness $h(t_i)$ at the present time $t_\circ$ of a wake created at time $t_i>t_{eq}$ is (Perivolaropoulos et al. 1990, Brandenberger 1991)
\begin{equation}
h(t_i)=\frac{24\pi}{5}G\mu v\gamma(v)z(t_i)^{1/2}t_\circ \ ,
\end{equation}
where $\gamma(v)$ is the relativistic factor associated with $v$.  The cosmic string model of structure formation is viable even if dark matter is hot since free streaming will not erase the non-adiabatic primordial perturbations on a small scale.  The wakes will have dimensions
\begin{equation}
v\xi(t_i)z(t_i)\times\xi(t_i)z(t_i)\times h(t_i) \ ,
\end{equation}
with mass
\begin{equation}
m(t_i)=\frac{24\pi}{5}G\mu v^2\gamma(v)\nu^{-2}\rho_\circ t_i^{1/3} t_\circ^{8/3} \ .
\end{equation}
The string correlation length is taken to be $\xi(t)=t/\nu$, where $\nu$ could in principle be determined from numerical simulations.  Hot dark matter wakes created before $t_{eq}$ are diluted by the radiation pressure and, therefore, do not contribute any significant amount to structure formation.  The string wake model predicts that planar wakes are the dominant structures in the large-scale galaxy distribution with the most numerous and thickest wakes created at $t_{eq}$.

In the string filament model, strings have a lot of small-scale structure and the string tension $|p|$ will be smaller than $\mu$.  The strings will have a small transverse velocity and will exert local gravitational attraction.  This induces a filamentary pattern of accretion.  Zanchin et al. (1996) have analyzed the string filament model for moderate transverse velocities and we will here present an extreme case in which the transverse velocity is negligible (Mitsouras et al. 1999).  Using the Zel'dovich approximation, the comoving radius $q(t_i)$ of a filament created at time $t_i$ is (Aguirre \& Brandenberger 1995)
\begin{equation}
q(t_i)=\left[\frac{12}{5}G\lambda\ln\left(\frac{t_\circ}{t_i}\right)\right]^{1/2}t_\circ \ ,
\end{equation}
where $\lambda=\mu-|p|$.  The mass in an individual filament is then
\begin{equation}
m(t_i)=\frac{12\pi}{5}G\lambda\nu^{-1}\rho_\circ t_i^{1/3}t_0^{8/3}\ln\left(\frac{t_\circ}{t_i}\right) \ ,
\end{equation}
where $\nu$ is defined as previously. 

In the toy topological defect simulations, galaxies are just laid down and no actual evolution takes place.  The procedure for setting up a string filament or string wake model first required dividing the time interval between $t_{eq}$ and $t_\circ$ into Hubble times.  At each time step $t_i$, a network of strings with correlation length $\xi(t_i)$ is laid down to simulate the evolution of the strings.  For each time step $t_i$, beginning with $t_{eq}$, we divided the simulation cube into $N(t_{i})$ Hubble volumes $t_i^{3}$.  A fixed number $c_s^{-3}$ of strings per Hubble volume is laid down, choosing string centers and directions at random in the cube.  Thus at each time step, $c_s^{-3}\times N(t_i)$ is the number of strings being laid down in the cube.  For given values of $G\lambda$ and $\nu$ in the case of the string filament simulation and $G\mu$, $\nu$ and $v$ in the case of the string wake simulation, the total number $c_s$ of strings is determined by demanding that the total mass satisfies critical density.  Galaxies are then laid down at random in the volume of the strings, with a uniform density amongst the strings.  For our string filament simulation, we have taken $G\lambda=10^{-6}$ to agree with the COBE normalization and $\nu=1$, such that the length of the string is equal to the Hubble radius at its time of creation.  The string wake simulation has values $G\mu=10^{-6}$ and $v\gamma=1$.

\subsection{Textures}

Textures are topological defects whose vacuum manifold $\cal M$ is a sphere $S^{3}$.  In 4-dimensional space-time, textures are unstable (Turok 1989) and can collapse to form seeds for structure formation.  While the energy in cosmic strings is trapped in potential energy, in textures, the energy is in spatial gradients with possibly some in kinetic energy.

The simplest texture configurations are spherically symmetric (Turok 1989).  At each time $t_i$, there is a finite probability $p$ that in a volume $t_i^{3}$, there will be a texture configuration entering the horizon (Prokopec 1991, Leese \& Prokopec 1991).  The initial texture configuration can be realized by a spherically symmetric velocity perturbation which results in a spherically symmetric density perturbation (Gooding et al. 1991).  The comoving radius $q(t_i)$ of this perturbation, calculated (Aguirre 1995) using the Zel'dovich approximation, is
\begin{equation}
q(t_i)=\frac{6}{5}\epsilon t_i^{-1/3}t_\circ^{4/3} \ ,
\end{equation}
where $\epsilon=G\eta^2$ and $\eta$ is the scale of symmetry breaking.

In the toy texture model, $p$ spherical texture configurations are laid down per Hubble volume $t_i^3$ with centers chosen at random in the box.  Galaxies are laid down within each texture with a Gaussian radial density profile and random angular distribution.  The number of galaxies per texture was chosen to satisfy critical density for the entire volume.  A value of $\epsilon=3\times10^5$ is an appropriate value for a COBE-normalized texture model (Pen et al. 1997).

\subsection{Construction of mock LCRS catalogs}

We construct mock catalogs by applying the LCRS selection criteria to the simulations.  The first step involved randomly assigning luminosities and corresponding absolute magnitudes to all the galaxies in each cube, according to the Schechter luminosity function
\begin{equation}
\phi(L)=\phi^*\left(\frac{L}{L^*}\right)^\alpha e^{-L/L^*} \ .
\end{equation}
where $\phi^*=0.019\ h^3 {\rm Mpc}^{-3}$, $\alpha=-0.70$, and $L^*$ corresponds to the absolute magnitude $M^*=-20.29+5\log h$ for the LCRS (Lin et al. 1996).  The faint end slope $\alpha=-0.70$ results from fitting the LCRS luminosity function and not from the absence of intrinsically faint galaxies in the survey.  The next step was to assign a surface brightness to the galaxies, but this presented a significant problem since the distribution of surface brightness is nontrivial.  While LSB galaxies are found in the same large scale structure as other galaxies, they tend to be less strongly clustered (Mo et al. 1994).  Since biasing may occur in one or more of the models, the LSB selection criterion has been omitted.

To apply the selection criteria, slices of appropriate geometry are first cut from the simulation cubes.  We have chosen to consider the three strips in the North galactic cap as a composite slice and the three strips in the South galactic cap as another composite slice.  The 3-dimensional geometry of the NGC and SGC slices should reveal more information about the topology of the galaxy distribution.  To avoid boundary problems in the 630 $h^{-1}$Mpc simulation cube, slices are cut from an interior cube with side-lengths of 490 $h^{-1}$Mpc.  After converting the absolute magnitudes into apparent magnitudes, the exact faint and bright apparent magnitude limits in each spectroscopic field (Shectman et al. 1996) are applied.  Next, the restriction on the maximum number of galaxies in each spectroscopic field is imposed, depending on the fiber system used for that particular field (random elimination of galaxies if the number of galaxies in the simulation field exceeds the maximal number of fibers used for that field).  Finally, to apply the minimum separation criterion without introducing bias, the galaxies are ordered in RA to form the set $\{g_i|g_1,...,g_N\}$.  Starting with galaxy $g_1$ and choosing to keep it, any galaxy within $55^{\prime\prime}$ is immediately removed from the set.  This step is iteratively carried out until the end of the set is reached.

\section{Minkowski Functionals}

Hadwiger's theorem states that in d spatial dimensions, the global morphological description of any spatial distribution can be completely characterized by d+1 descriptors comprising the set of Minkowski functionals (Hadwiger 1957).  Given an isodensity surface $\partial K$ in 3d, the morphology is completely determined by the following four statistics.  The first Minkowski functional
\begin{equation}
M_\circ=V \ ,
\end{equation}
is simply the volume enclosed by the surface $\partial K$ and the second functional
\begin{equation}
M_1=\frac{A}{8} \ ,
\end{equation}
is related to the surface area $A$.  The third functional
\begin{equation}
M_2=\frac{H}{2\pi_2} \ ,
\end{equation}
provides information about shape and is proportional to the integrated mean curvature
\begin{equation}
H=\frac{1}{2}\int_{\partial K}\left(\frac{1}{R_1}+\frac{1}{R_2}\right)dA \ ,
\end{equation}
where $R_1$ and $R_2$ are the principal radii of the surface $\partial K$.  The fourth Minkowski functional
\begin{equation}
M_3=\frac{3\chi}{4\pi} \ ,
\end{equation}
which is related to the Euler characteristic
\begin{equation}
\chi=\int_{\partial K}\frac{1}{R_1R_2}dA \ .
\end{equation}
is purely a topological quantity and can be geometrically defined as
\begin{equation}
\chi\propto{\rm (\#\ of\ clusters)}-{\rm (\#\ of\ tunnels)}+{\rm (\#\ of\ cavities)} \ .
\end{equation}

The Minkowski functionals are realized for a galaxy distribution by placing spheres of radius $r$ centered on the galaxies.  The union of these spheres specifies an ensemble of clusters, where two spheres are said to belong to the same cluster if they are connected by a chain of intersecting spheres.  The Minkowski functionals will be used to measure the topology of the coverage.  We use the method of Mecke et al. (1994) and Kerscher et al. (1997) and display the functionals as a function of the length scale r, which acts as a smoothing length of the density field.

\section{Results}

For each model studied, the simulations were performed in the following way. First, a cubical box of comoving length 630 $h^{-1}$Mpc was populated with galaxies according to the prescriptions discussed in Section 3. Next, slices with the appropriate geometry of the LCRS survey were cut out from the simulation cube. To avoid boundary effects, the slices were extracted from an interior cube with side length of 490 $h^{-1}$Mpc. For comparing with the individual LCRS slices, four galaxies in the interior box were chosen as centers of slightly oversized slices of $90^\circ$ in length and $1.6^\circ$ in width. For comparing with the composite LCRS slices, four larger slices of $90^\circ$ in length and $10.6^\circ$ in width were cut out.

Luminosities were assigned randomly in the initial simulation box according to the Schechter luminosity function. For each slice, this then allows us to determine the apparent magnitude of all galaxies contained in the slice. Next, the LCRS selection criteria were applied as detailed in Section 3.4.

For each slice, the Minkowski functionals were calculated using a slightly modified version (Mitsouras 1998) of the program package publically available (Kerscher et al. 1998). Averages over the four centers were taken. Under the ergodic hypothesis, the volume averages (averages over the slices) will be equal to the ensemble averages over different realizations of the theories. The error bars indicated in our plots correspond to these statistical errors.

Figures 1 - 4 show the resulting volume-averaged Minkowski functionals for the NGC slice. The results of our simulations are compared with the LCRS data. As mentioned in the Introduction, none of the models considered in this paper is able to explain the data. The SCDM model gives the least bad fit, but even it is ruled out with high statistical significance. \footnote{The oscillations seen at large values of $r$ are artifacts of the boundary conditions.}

As is apparent from Figures 1 - 4, the Minkowski functionals provide a very powerful tool to distinguish between models with identical power spectra but different topology. Amongst the Minkowski functionals, the integrated mean curvature statistic M2 appears to be the most powerful discriminant, followed by the genus statistic M3. This confirms the results seen in previous studies (Mitsouras 1998, Mitsouras et al. 1999, Kerscher et al. 1997).

The predictions of the texture toy model differ dramatically from those of the other models studied, and also from the data. The reason is that in the texture toy model, structure is dominated by well separated spherical clumps, whereas in the other models there is a higher degree of connectedness of the individual structures. In addition, the central density in the texture toy model is larger than in the other theories studied. Therefore, M0, the volume enclosed by the ``isodensity" surface, is significantly smaller for any value of $r$ than in the other models. The surface area (M1) itself continues to increase with $r$ since to a first approximation the number of disconnected components remain the same while the size of each object increases. In the other models studied, the value of M1 for small $r$ is higher since the shape of the objects does not minimize the surface area for fixed enclosed volume. However, as $r$ increases, the growing interconnectedness of the components leads, for large $r$, to a gradual decrease in the value of M1. The radius corresponding to the maximum of M1 corresponds to the mean separation between clusters.

For a single component surface without holes and with fixed volume, the sphere minimizes the integrated mean curvature. Hence, for small values of $r$, the value of M2 is lower in the texture model than in the other models. In particular, in the string filament and wake toy models the individual filaments and wakes have a small curvature radius in one direction, and hence a large mean curvature. Once the individual filaments and wakes merge by overlapping, the overall shape of the surface becomes concave and hence M2 turns negative. This feature is shared by the SCDM model, and by the data. Similarly, the overlapping of filaments and wakes leads to a structure with many holes for larger values of $r$. This leads to negative genus for these scales, a feature which is again shared by the SCDM model and by the data, but is absent in the texture toy model. The radius corresponding to the minimum of M3 gives another good indicator of the mean separation of clusters.  

By comparing the LCRS data with the simulations, it appears that the data has less connectivity than what would be seen in the SCDM model, and significantly less than what would be predicted by the toy string theories. These results are confirmed by the analysis of the SGC slice (Figures 5 - 8). Note that the differences between the NGC and SGC slices are due to the different selection effects.

The number of galaxies in the model slices is not the same as in the LCRS slices. In order to check that the disagreement between the theories and the data is not due to this difference, we have repeated the analysis with ``randomly sampled" model slices, model slices from which in a random way galaxies were removed in order to obtain the same number of galaxies as in the actual LCRS slices. Figures 9 - 12 give the results for the ``randomly sampled" NGC slice, Figures 13 - 16 the corresponding results for the SGC slice. As is apparent, the conclusions are unchanged.

As mentioned in Section 3.4, we did not apply the LSB selection criterion of the observational survey. To investigate whether this omission could account for the disagreement between the data and the SCDM model, we repeated the calculation of the Minkowski functionals of the observational data with the lowest $20\%$ of the galaxies (in terms of surface brightness) removed. There were no appreciable changes in the results.

\section{Discussion}

The main conclusion of our study is that Minkowski functionals provide a very powerful statistical tool to differentiate between theories with identical power spectra but different topologies. We have compared the predictions of the SCDM model and of three topological defect toy models with the LCRS data, taking careful account of most of the selection criteria of the LCRS survey. 

None of the models studied comes close to giving a satisfactory match to the data. This is not surprising since the SCDM model is known not to correctly describe the observed clustering of matter in the Universe, and since the defect models considered here are all unrealistic toy models. The texture model produces isolated and well-separated dense clusters and gives a picture of the distribution of galaxies which already at first glance is inconsistent with any observational data set. The string models are based on ideal straight filaments or wakes, and the filaments and walls seen in observations are not straight.

It would be of great interest to explore whether more realistic models can provide a better fit to the data. In the class of inflation-based models yielding a spectrum of roughly scale-invariant primordial perturbations it would be of interest to explore the effects of reducing $\Omega_m$ and introducing a cosmological constant $\Lambda$. In particular, it would be of great interest to see if the Minkowski functionals as applied to the LCRS geometry can differentiate between the set of inflation-based models which are compatible with the present CMB data and with the amplitude of the matter power spectrum on large scales. Work on this issue is in progress (Merali et al. 2000).

From the point of view of particle cosmology, the defect models studied in this paper are too idealized to be realistic. In texture models the resulting structures are not all spherically symmetric, and there are important non-topological fluctuations which also lead to density inhomogeneities (Borrill 1994). Similarly, the structure seeded in a cosmic string model will also include effects of string loops (see e.g. Brandenberger 1994 for a review), and the structures seeded by the long string network will be given by a combination of the filament and wake effects. In addition, the structures will not be straight. It would also be of interest to use the Minkowski functionals to study the predictions of numerically simulated more realistic defect models, and to see if there are parameters for which good agreement with the data is obtained.

It is also possible that the differences between the theoretical predictions and the LCRS data are due to biasing effects. This issue could be investigated by using hydrodynamical simulations of structure formation.  Redshift distortions also complicate the interpretation of clustering in redshift surveys.  The peculiar velocities of galaxies cause them to appear displaced along the line of sight in redshift space, thus causing the observed galaxy distribution to be anisotropic.  It would be interesting to compare the real and redshift space results from N-body simulations using Minkowski functionals.  

In conclusion, we hope to have convinced the reader that Minkowski functionals provide a very powerful tool for quantitative cosmology.

\newpage
\begin{figure}
\epsfxsize=3.3in
\epsfbox{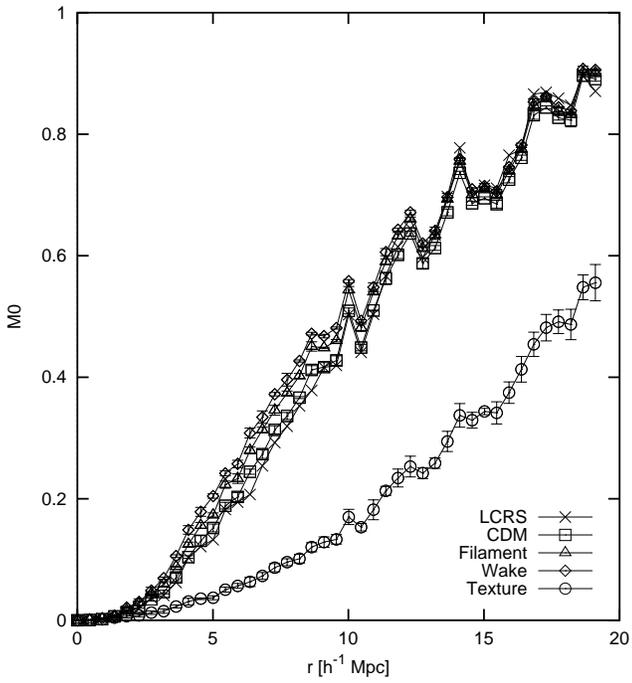}
\caption{M0 NGC slice without random sampling.}
\end{figure}

\begin{figure}
\epsfxsize=3.3in
\epsfbox{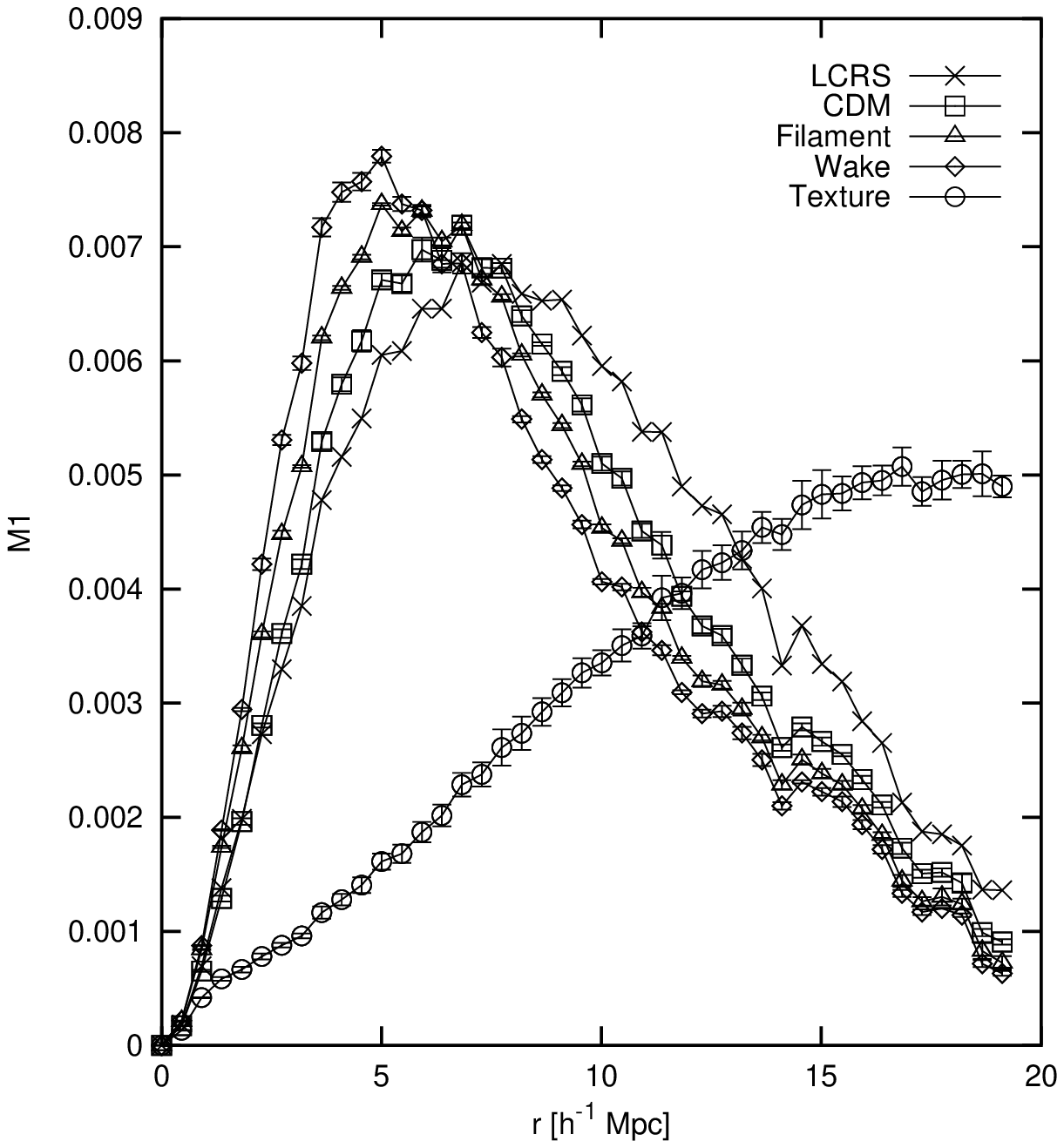}
\caption{}
\end{figure}

\begin{figure}
\epsfxsize=3.3in
\epsfbox{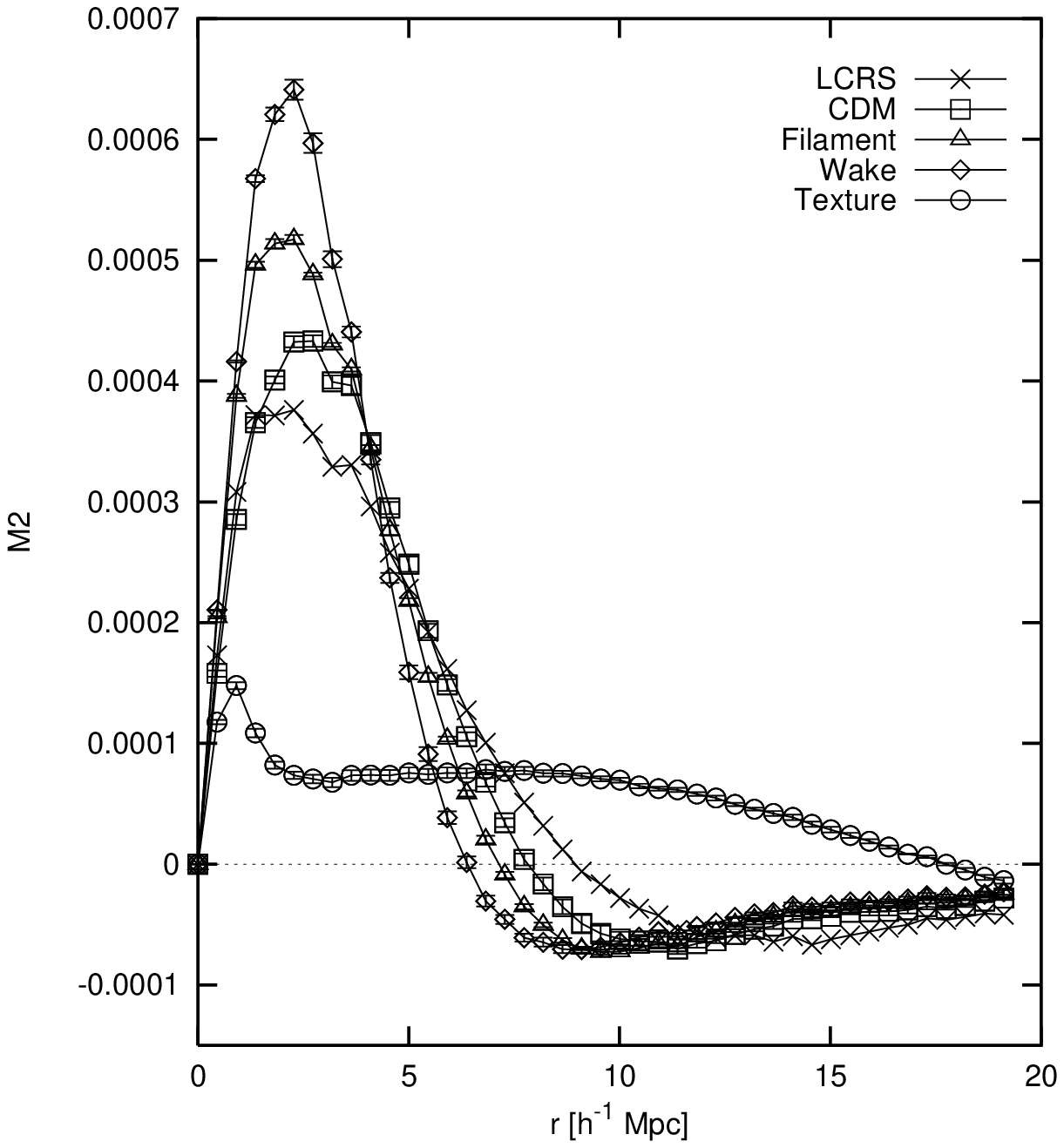}
\caption{}
\end{figure}

\begin{figure}
\epsfxsize=3.3in
\epsfbox{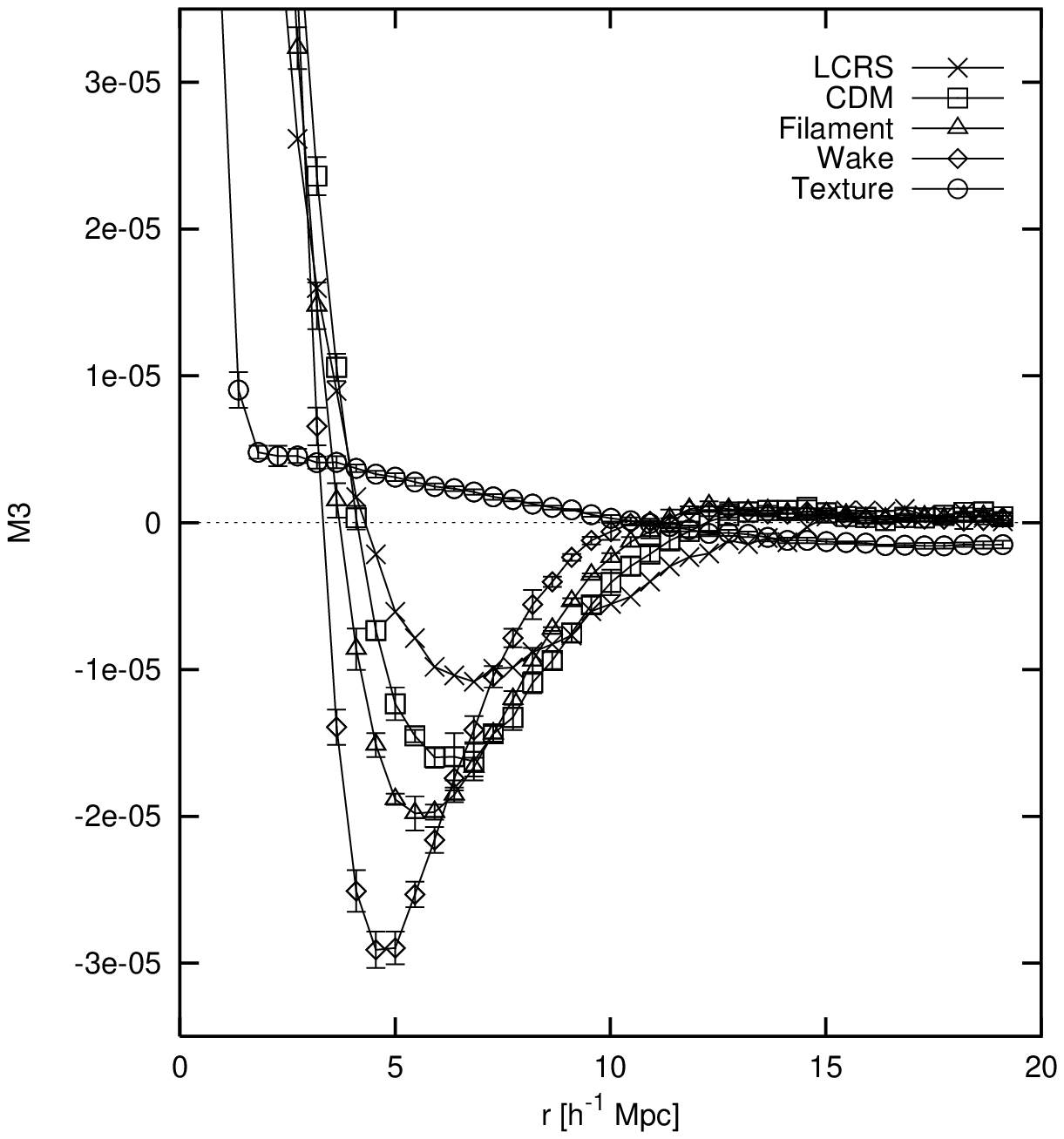}
\caption{}
\end{figure}

\newpage
\begin{figure}
\epsfxsize=3.3in
\epsfbox{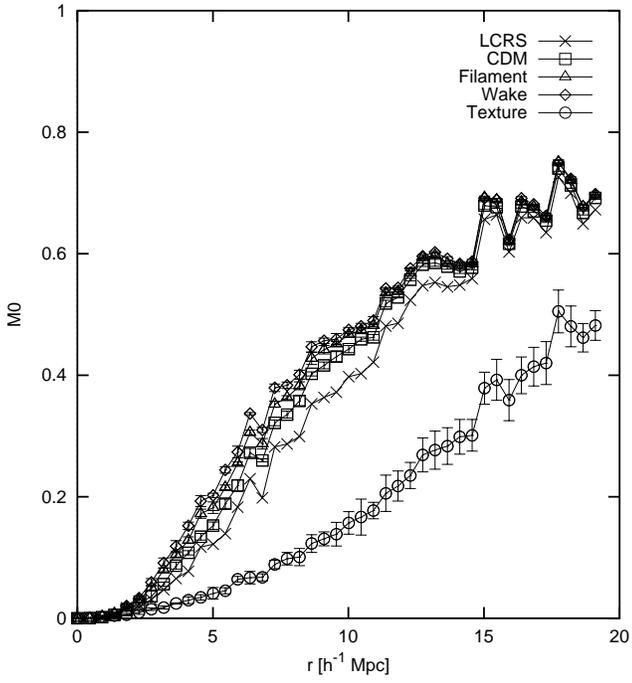}
\caption{SGC slice without random sampling.}
\end{figure}

\begin{figure}
\epsfxsize=3.3in
\epsfbox{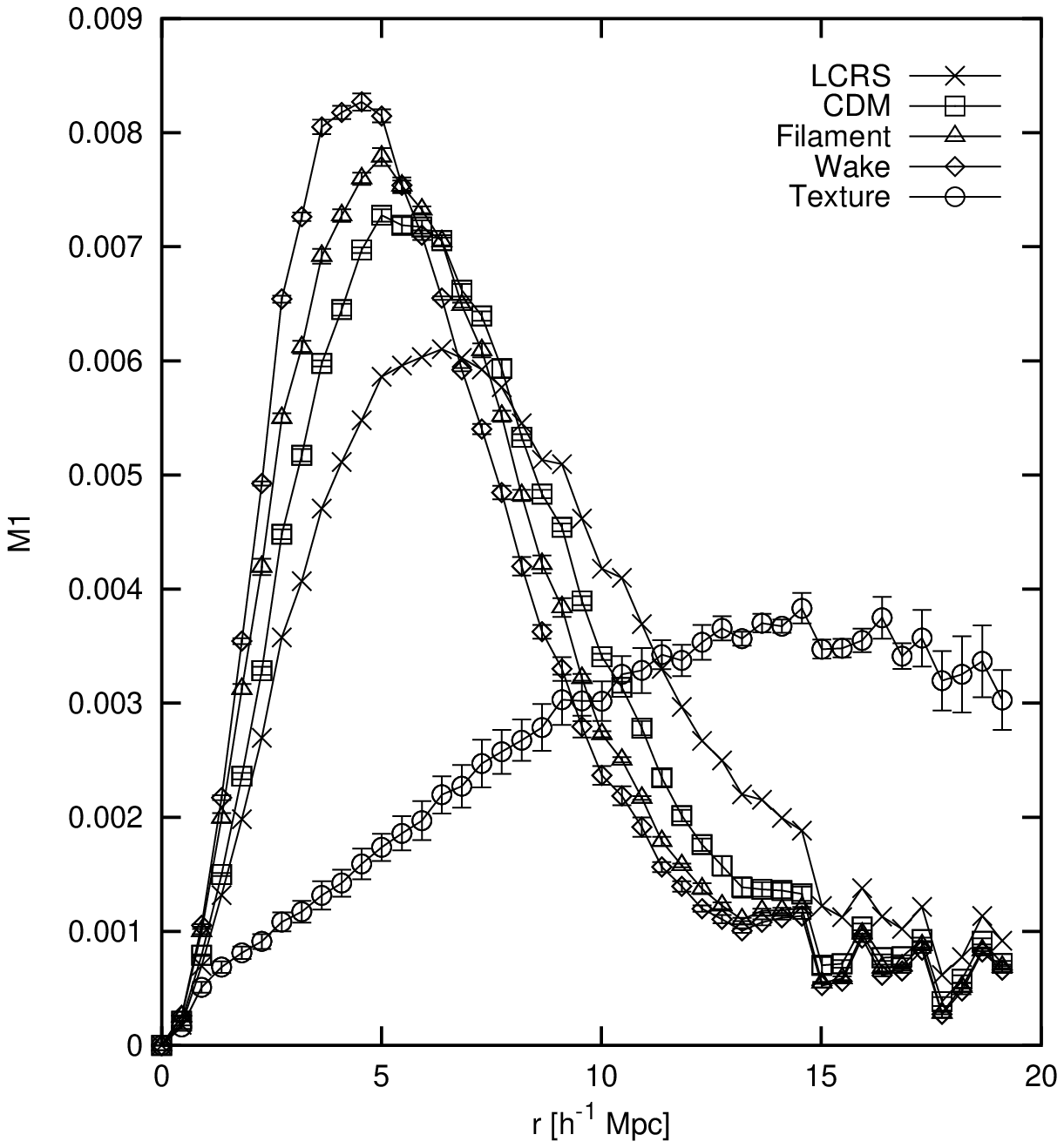}
\caption{}
\end{figure}

\begin{figure}
\epsfxsize=3.3in
\epsfbox{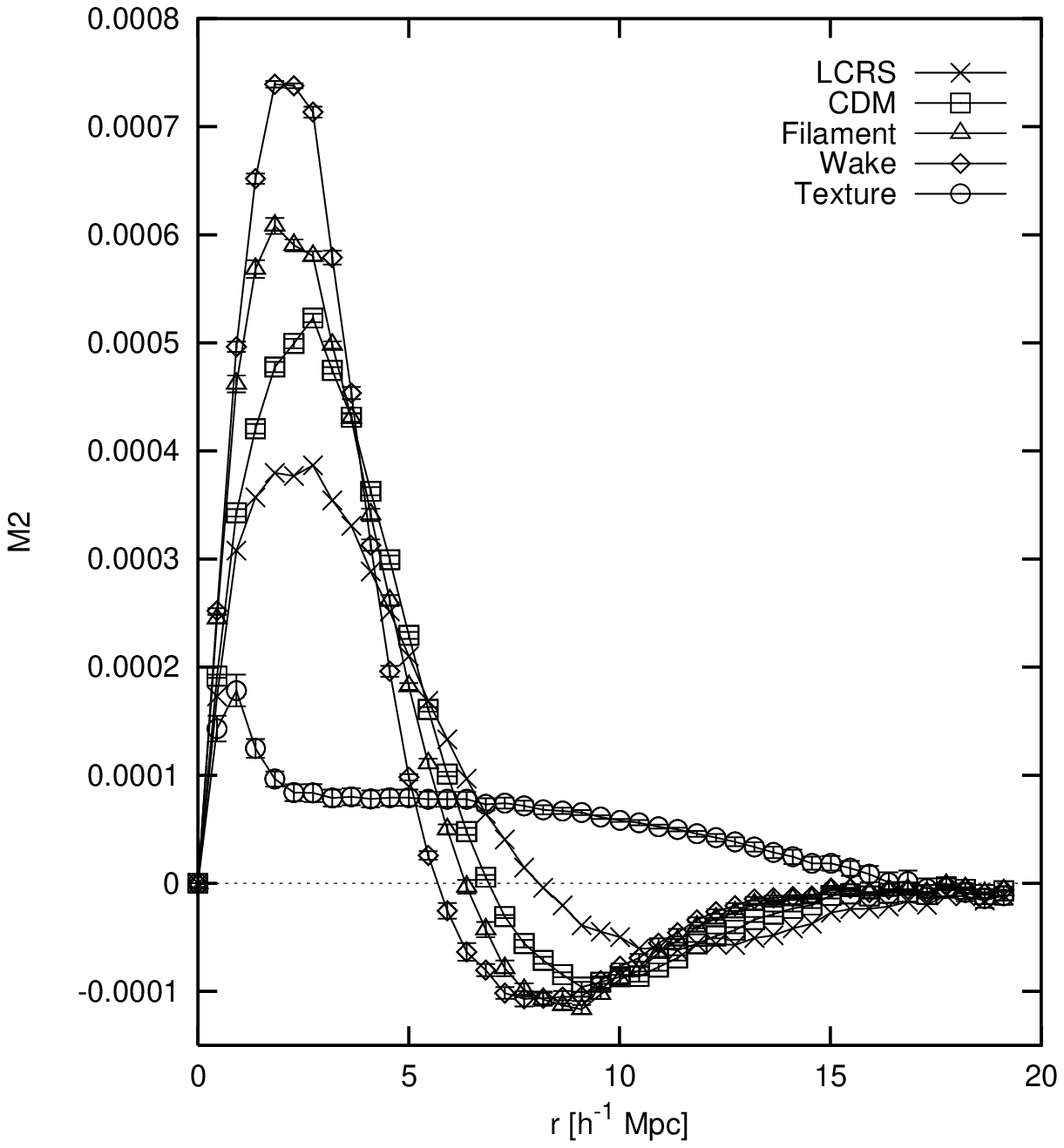}
\caption{}
\end{figure}

\begin{figure}
\epsfxsize=3.3in
\epsfbox{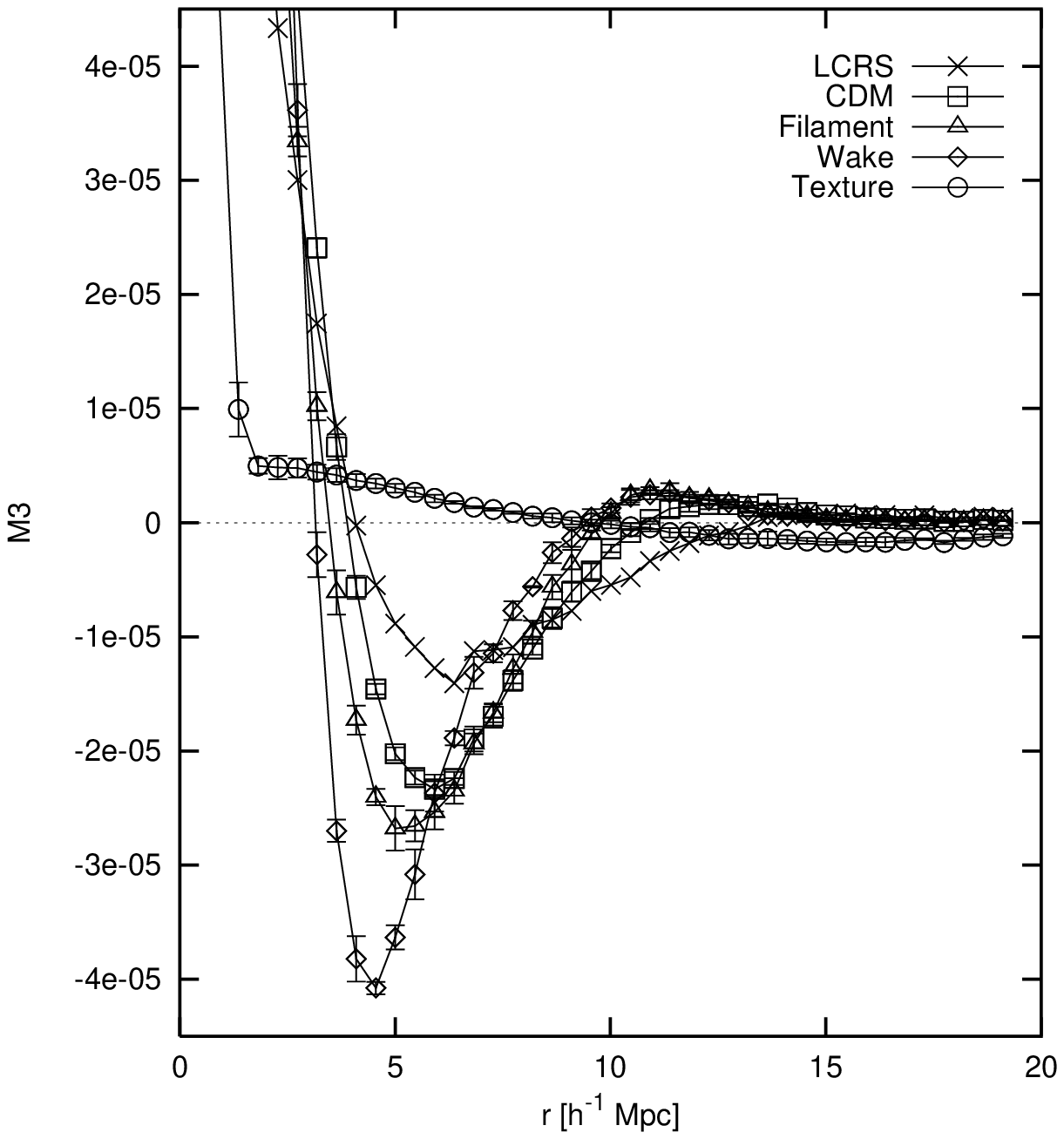}
\caption{}
\end{figure}

\newpage
\begin{figure}
\epsfxsize=3.3in
\epsfbox{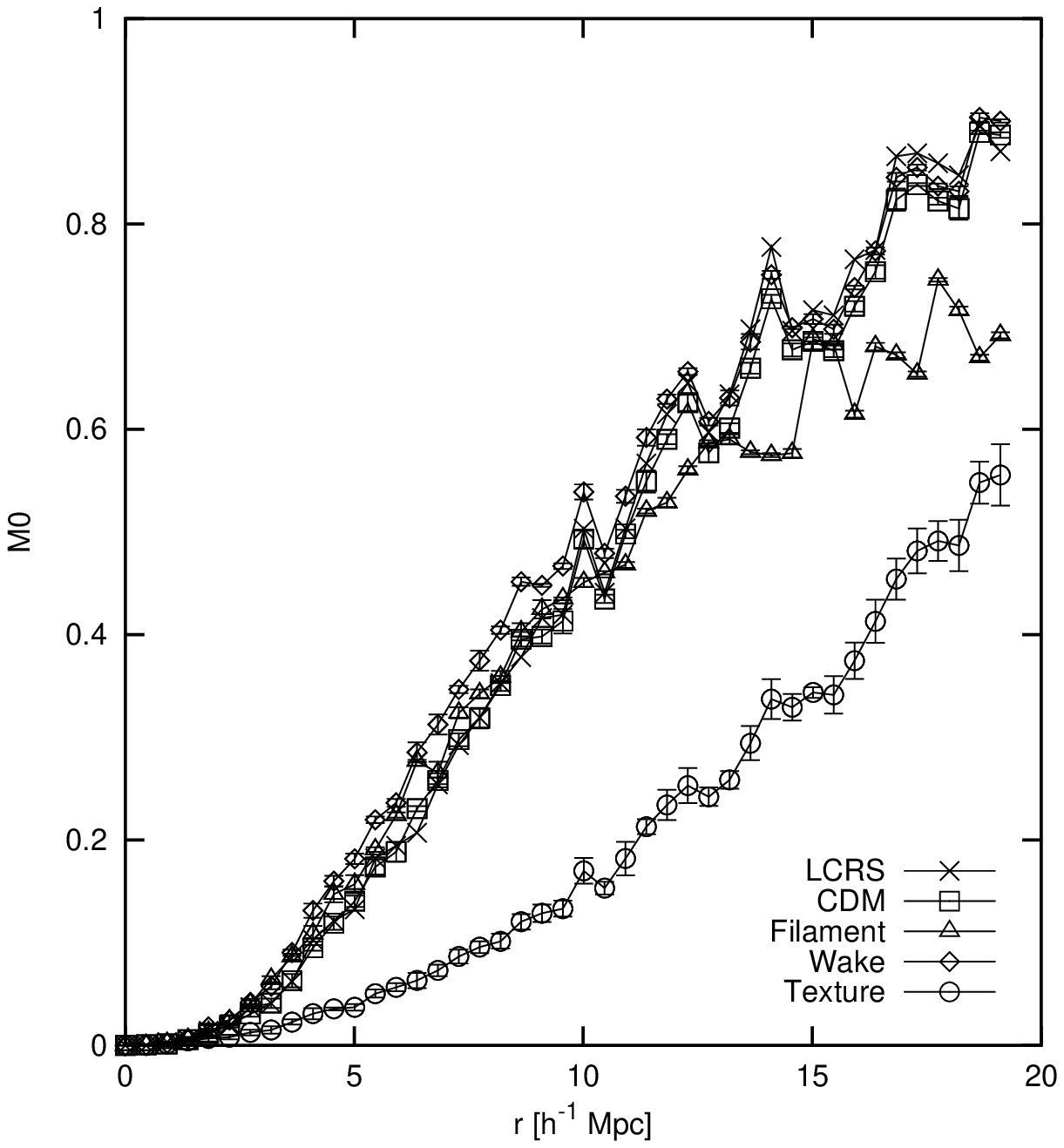}
\caption{NGC slice normalized with respect to the average number of galaxies in the LCRS NGC slice.}
\end{figure}

\begin{figure}
\epsfxsize=3.3in
\epsfbox{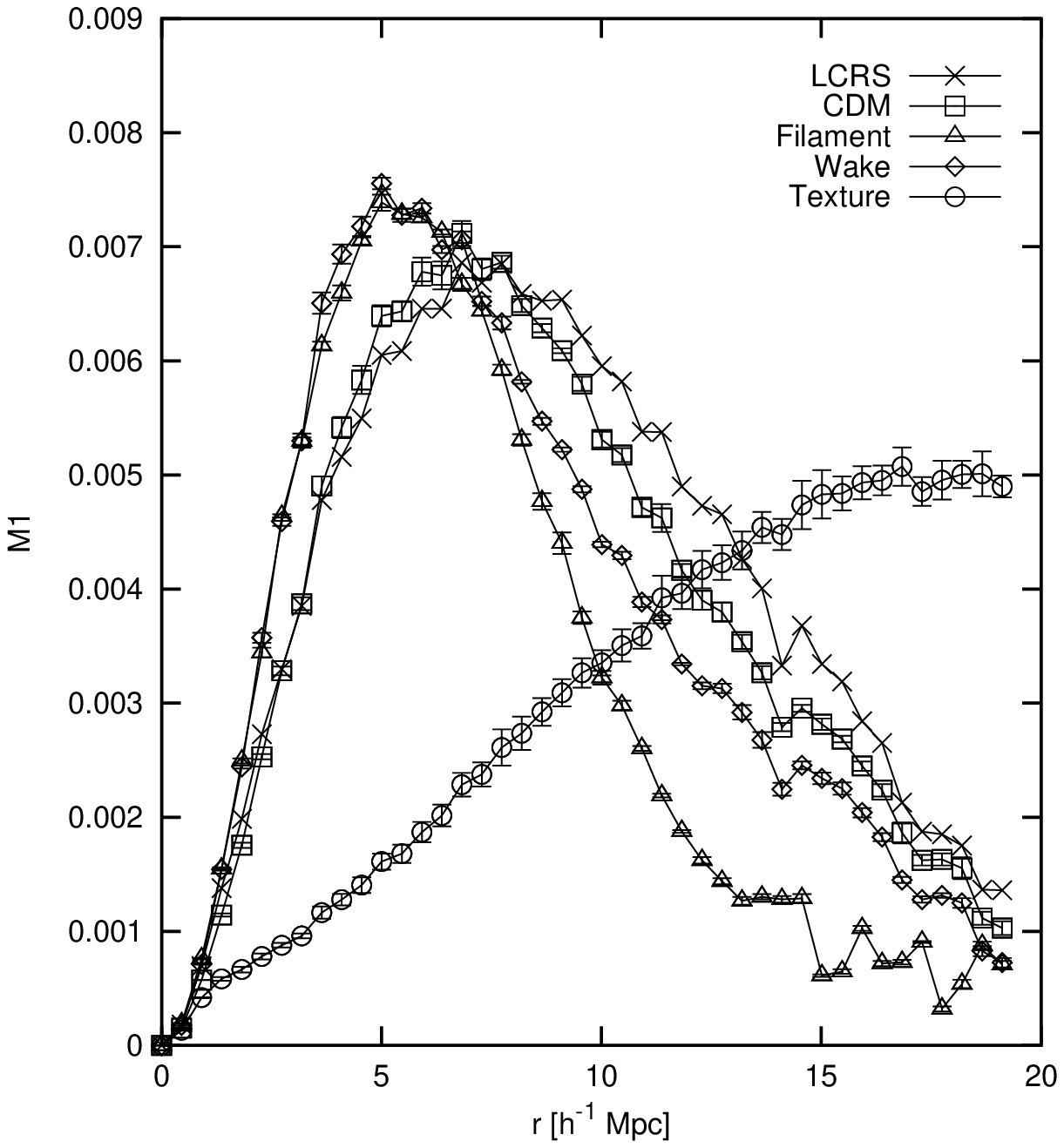}
\caption{}
\end{figure}

\begin{figure}
\epsfxsize=3.3in
\epsfbox{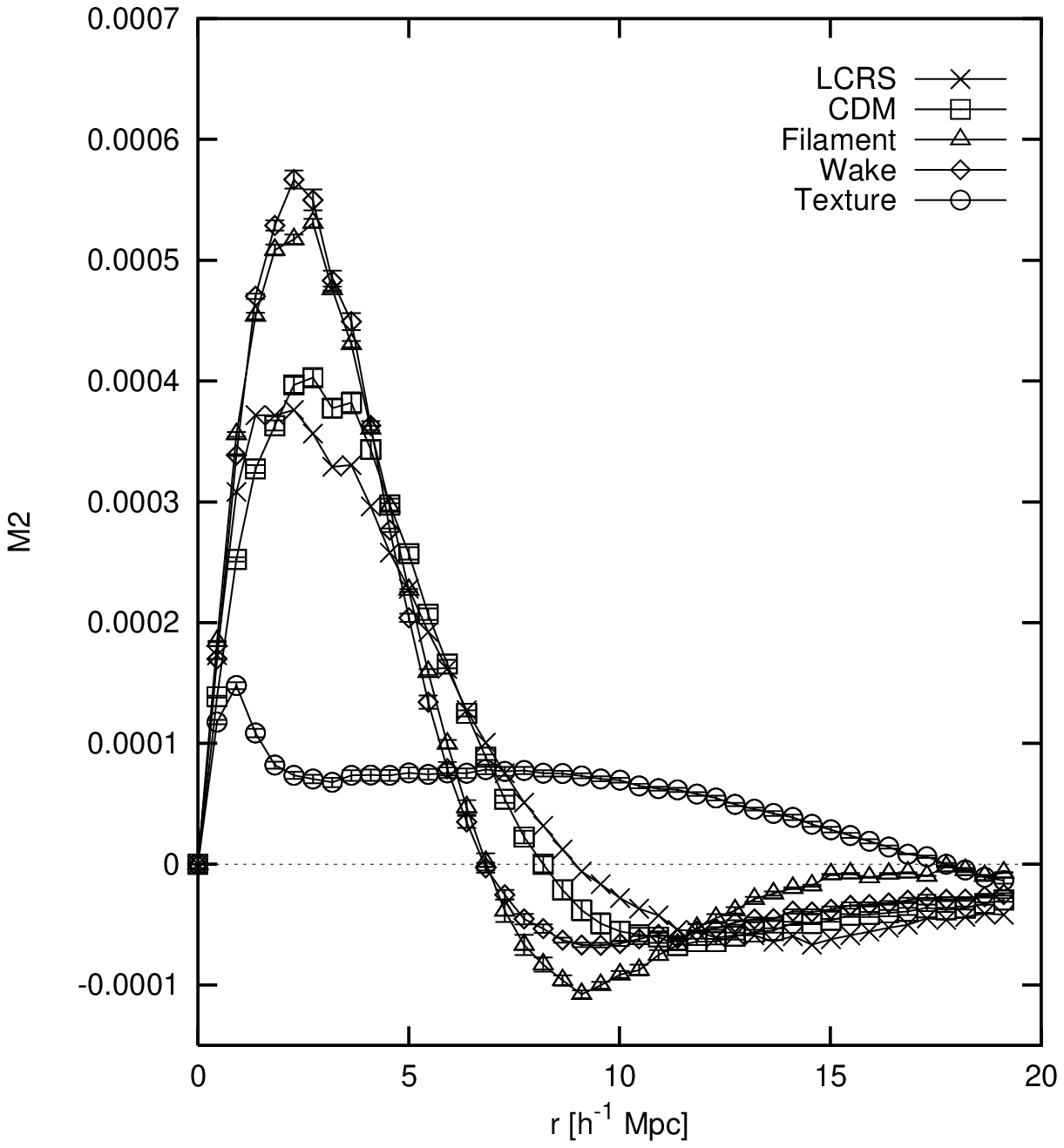}
\caption{}
\end{figure}

\begin{figure}
\epsfxsize=3.3in
\epsfbox{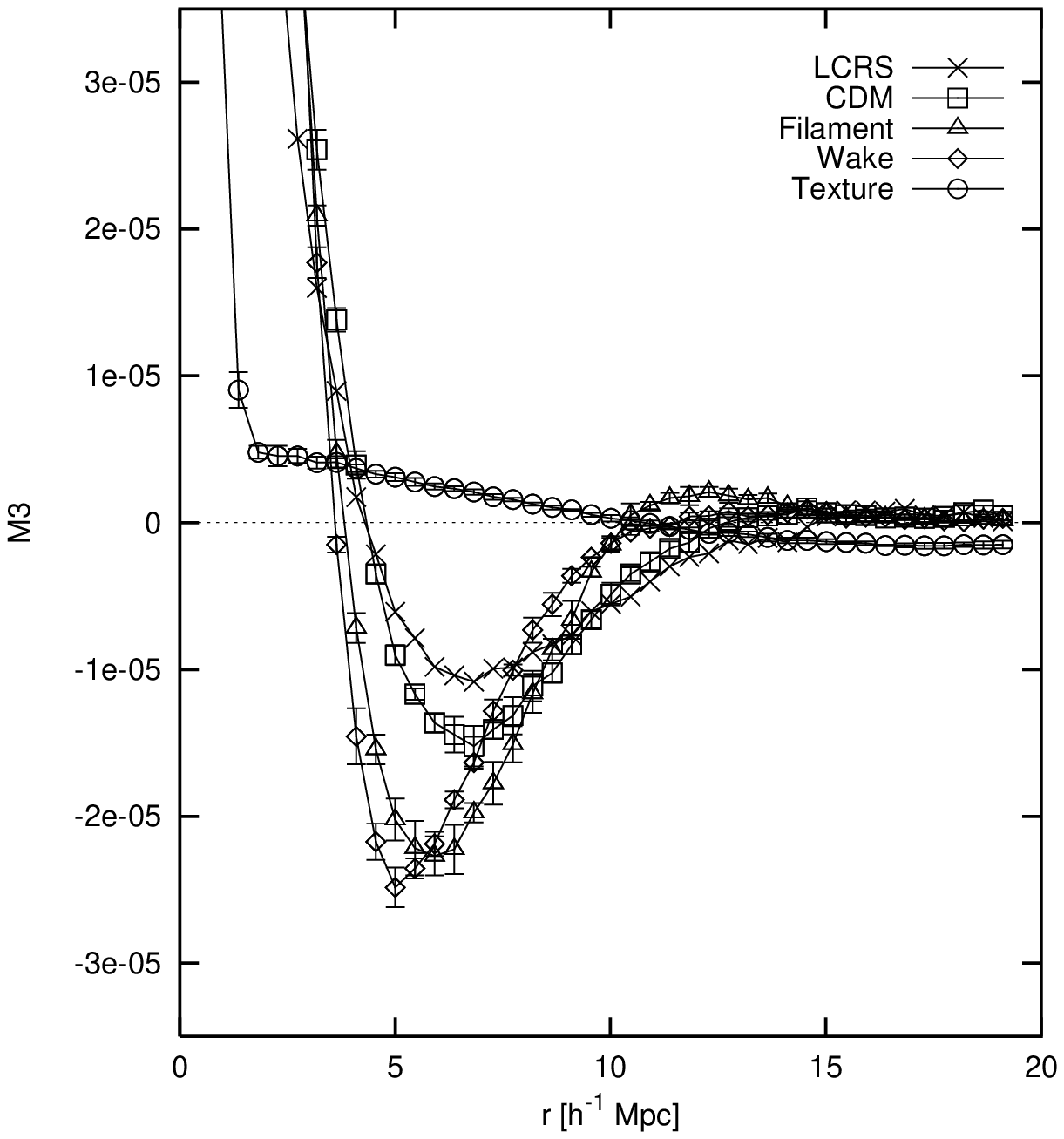}
\caption{}
\end{figure}

\newpage
\begin{figure}
\epsfxsize=3.3in
\epsfbox{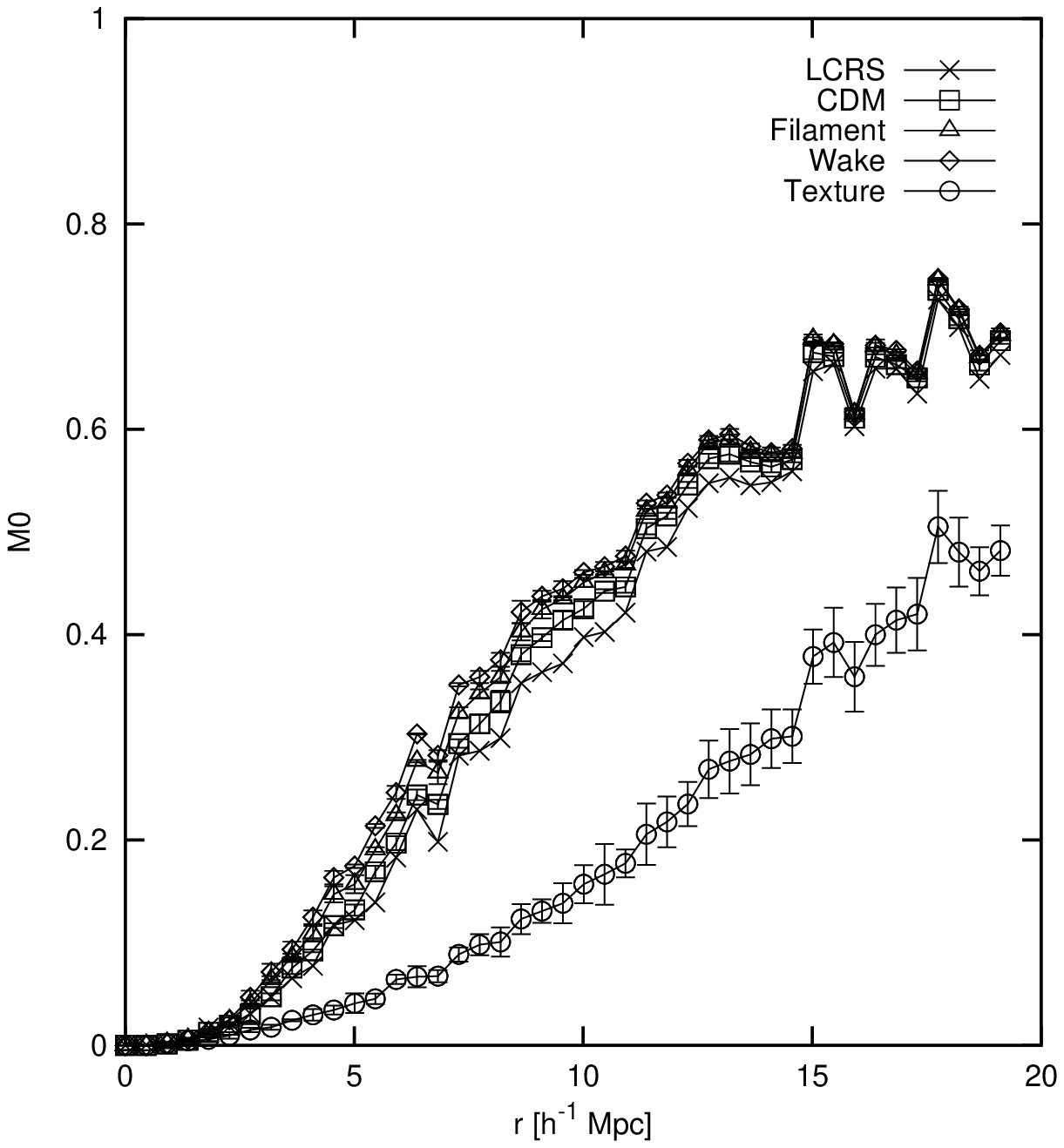}
\caption{SGC slice normalized with respect to the average number of galaxies in the LCRS SGC slice.}
\end{figure}

\begin{figure}
\epsfxsize=3.3in
\epsfbox{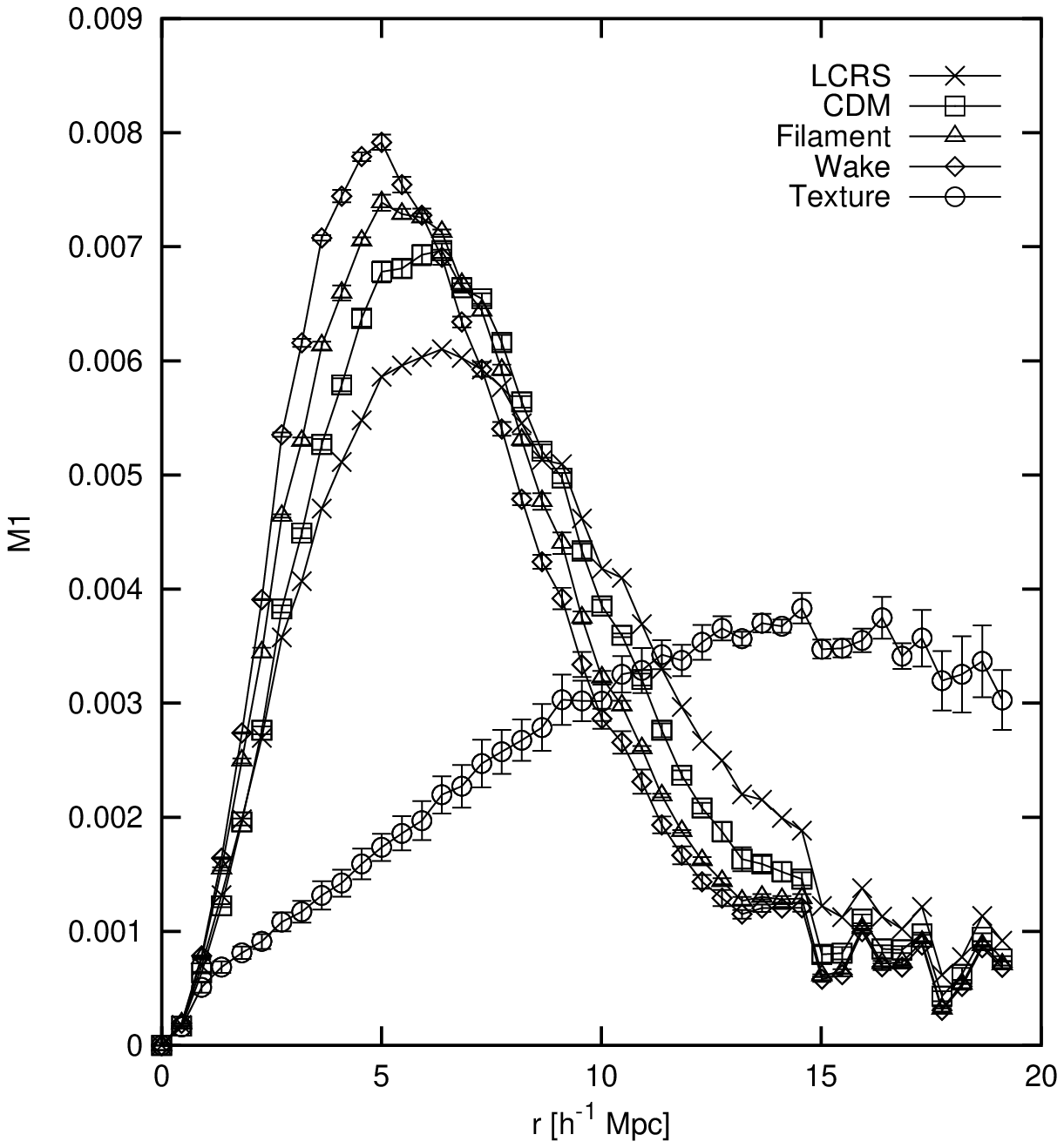}
\caption{}
\end{figure}

\begin{figure}
\epsfxsize=3.3in
\epsfbox{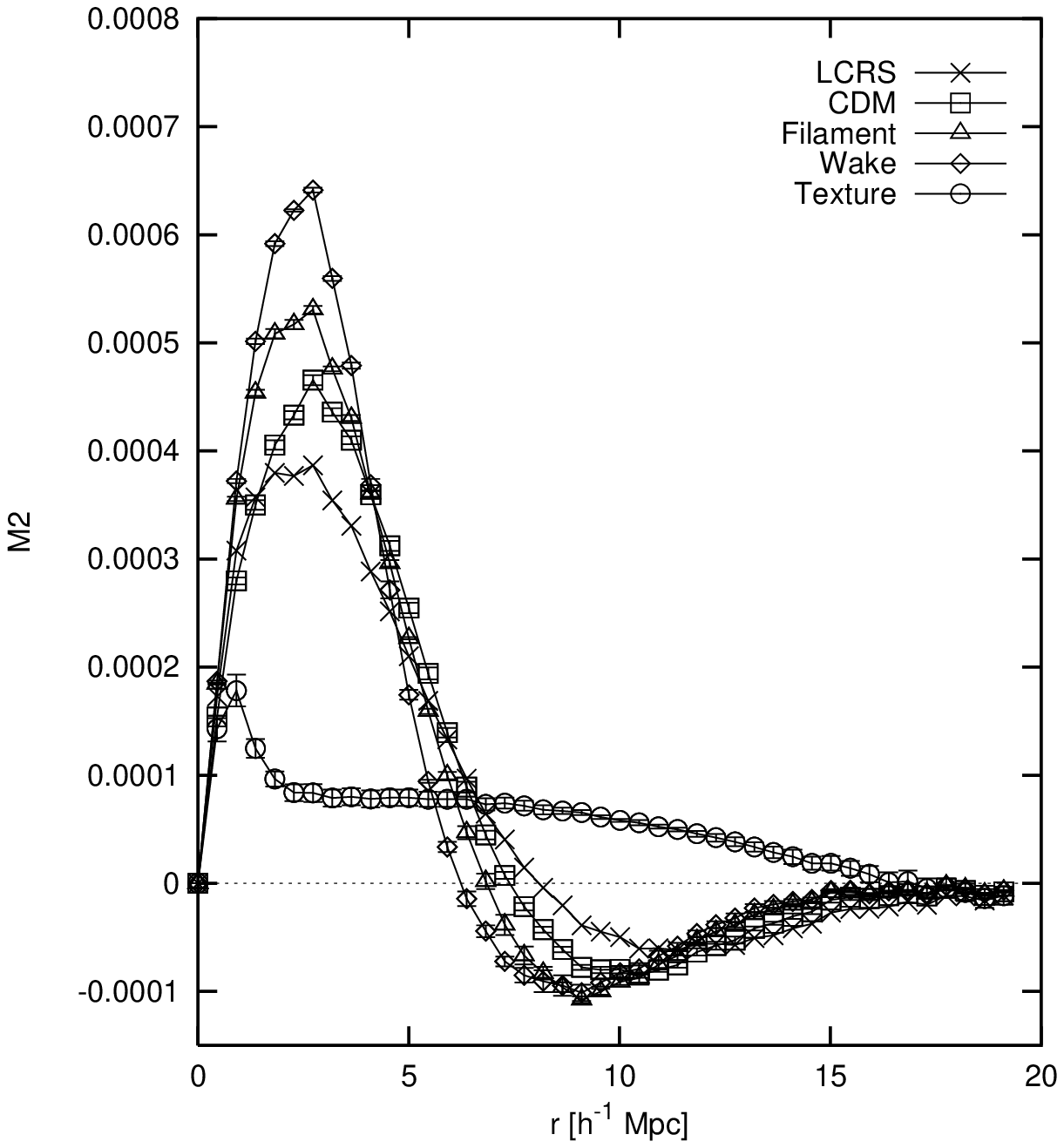}
\caption{}
\end{figure}

\begin{figure}
\epsfxsize=3.3in
\epsfbox{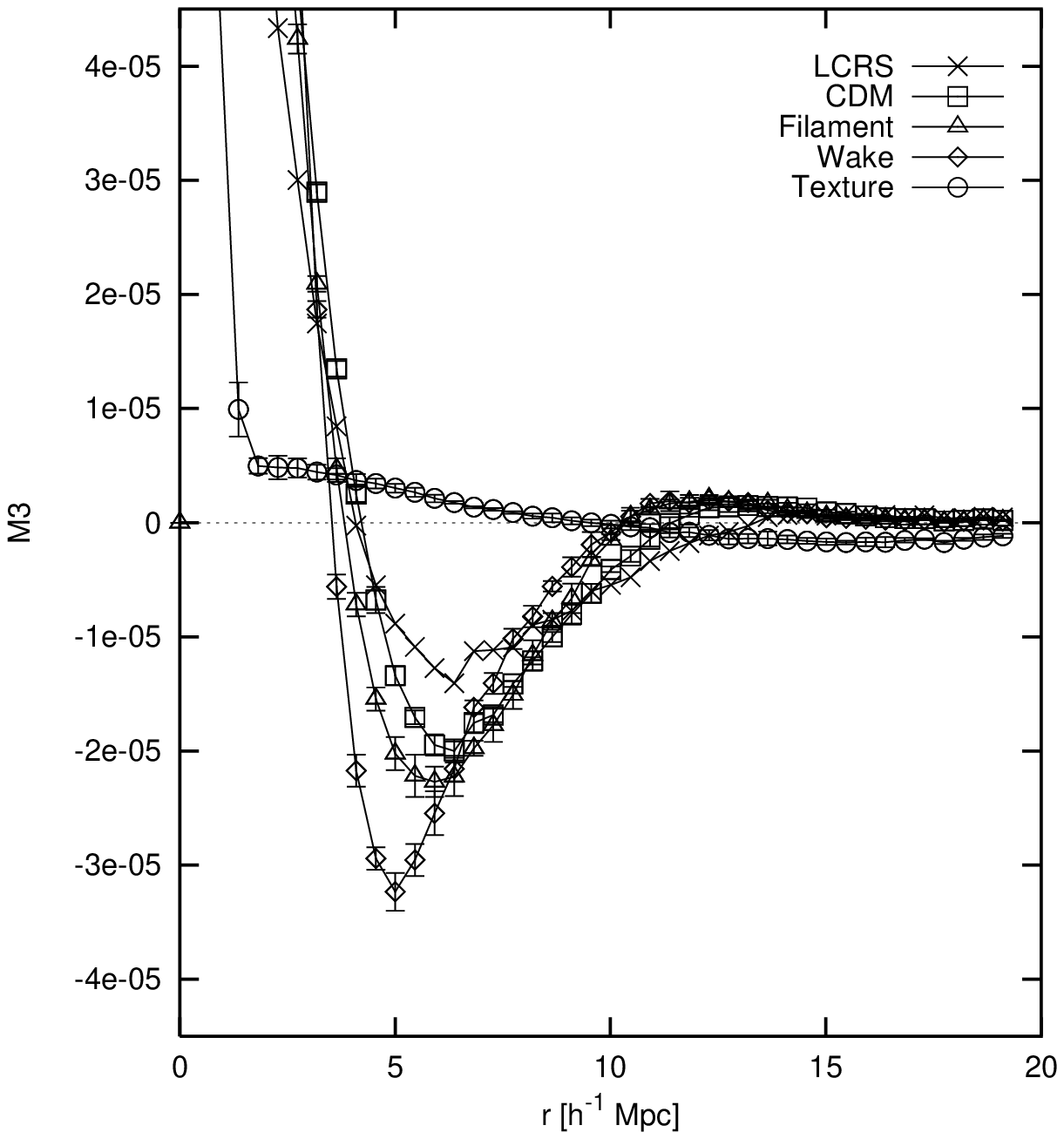}
\caption{}
\end{figure}


\begin{thebibliography}{}
\bibitem{} Aguirre, A., 1995, Senior thesis, Brown Univ. (unpublished).
\bibitem{} Aguirre, A. \& Brandenberger, R., 1995, Int. J. Mod. Phys., D4, 711; astro-ph/9505031.
\bibitem{} Albrecht, A., Battye, R. \& Robinson, J., 1999, Phys. Rev. D59, 023508.
\bibitem{} Babul, A. \& Starkman, G., 1992, Ap. J. 401, 28.
\bibitem{} Bertschinger, E. \& Gelb, J., 1991, Computers in Physics, 5, 164.
\bibitem{} Bharadwaj, S., Sahni, V., Sathyaprakash, B., Shandarin, S. \& Yess, C., 2000, Ap. J. 528, 21.
\bibitem{} Borrill, J., 1994, Phys. Rev. D50, 3676.
\bibitem{} Bouchet, R., Peter, P., Riazuelo, A. \& Sakellariadou, M., 2000, astro-ph/0005022.
\bibitem{} Brandenberger, R., 1991, Phys. Scripta, T36, 114.
\bibitem{} Brandenberger, R., 1994, Int. J. Mod. Phys. A9, 2117.
\bibitem{} Chibisov, G. \& Mukhanov, V., 1980, `Galaxy Formation and Phonons,' Lebedev Physical Institute Preprint, No. 162.
\bibitem{} Chibisov, G. \& Mukhanov, V., 1982, MNRAS, 200, 535.
\bibitem{} Colley, W., Gott, J. R., Weinberg, D., Park, C. \& Berlind, A.. 1999,
astro-ph/9902332.
\bibitem{} de Bernardis, P. et al., 2000, Nature 404, 955.
\bibitem{} de Lapparent, V., Geller, M. \& Huchra, J., 1986, Ap. J. (Lett.) 302, L1. 
\bibitem{} Doroshkevich, A. et al.. 1996, MNRAS 283, 128.
\bibitem{} Gooding, A., Spergel, D. \& Turok, N., 1991, Ap. J. (Lett.), 372, L5.
\bibitem{} Hadwiger, H., 1957, Vorlesungen \"{u}ber Inhalt, Oberfl\"{a}che und Isoperimetrie, Springer.
\bibitem{} Hanany, S. et al., 2000, astro-ph/0005123.
\bibitem{} Haynes, M. \& Giovanelli, R., 1986, Ap. J. (Lett.), 306, L55.
\bibitem{} Kerscher, M. et al., 1997, MNRAS, 284, 73; astro-ph/9606133.
\bibitem{} Kerscher, M., Schmalzing, J., Buchert, T. \& Wagner, H.. 1998, Astron. Astrophys. 333, 1.
\bibitem{} Kibble, T. W. B., 1976, J. Phys., A9, 1387.
\bibitem{} Kirshner, R., Oemler, A., Schechter, P. \& Shectman, S., 1981, Ap. J. (Lett.), 248, L57.
\bibitem{} Leese, R. \& Prokopec, T., 1991, Phys. Rev., D44, 3749.
\bibitem{} Lin, H., Kirshner, R. P., Shectman, S. A., Landy, S. D., Oemler, A., Tucker, D. L., \& Schechter, P. L., 1996, ApJ, 464, 60.
\bibitem{} Lukash, V., 1980, Pis'ma Zh. Eksp. Teor. Fiz., 31, 631.
\bibitem{} Mecke, K. R., Buchert, T., \& Wagner, H., 1994, Astron. Astrophys., 288, 697.
\bibitem{} Merali, Z., Trac, H., G\"otz, M. \& Brandenberger, R., in preparation.
\bibitem{} Mitsouras, D., 1998, Senior Thesis, Brown Univ. (unpublished).
\bibitem{} Mitsouras, D., Brandenberger, R., \& Hickson, P., 1999, MNRAS, 305, 19; astro-ph/9806360.
\bibitem{} Mo, H., McGaugh, S. S., \& Bothun, G. D., 1994, MNRAS, 267, 129; astro-ph/9311004.
\bibitem{} Moessner, R., Perivolaropoulos, L. \& Brandenberger, R.. 1994, Ap. J. 425, 365.
\bibitem{} Mukhanov, V. \& Chibisov, G., 1981, JETP (Lett.) 33, 532.
\bibitem{} Mukhanov, V., Feldman, H \& Brandenberger, R.. 1992, Phys. Rep. 215, 203.
\bibitem{} Pen, U.-L., 1998, Ap. J. 498, 60.
\bibitem {}Pen, U.-L., Seljak, U. \& Turok, N., 1997, Phys. Rev. Lett. 79, 1611.
\bibitem{} Perivolarapoulos, L., Brandenberger, R., \& Stebbins, A., 1990, Phys. Rev., D41, 1764.
\bibitem{} Press, W., 1980, Phys. Scr., 21, 702. 
\bibitem{} Prokopec, T., 1991, Phys. Lett., 262B, 215.
\bibitem{} Sahni, V., Sathyaprakash, B. \& Shandarin, S., 1998, Ap. J. (Lett.), 495, L5.
\bibitem{} Sathyaprakash, B., Sahni, V. \& Shandarin, S., 1996, Ap. J. (Lett.),
462, L5.
\bibitem{} Sathyaprakash, B., Sahni, V., Shandarin, S. \& Fisher, K., 1998, Ap. J. (Lett.), 507, L109.
\bibitem{} Sato, K., 1981, MNRAS 195, 467.
\bibitem{} Schmalzing, J. \& Buchert, T.. 1997, Ap. J. 482, 1.
\bibitem{} Schmalzing, J., Buchert, T., Melott, A., Sahni, V., Sathyaprakash, B. \& Shandarin, S.. 1999, Ap. J. 526, 568.
\bibitem{} Shectman, S. A., Landy, S. D., Oemler, A., Tucker, D. L., Lin, H., Kirshner, R. P., \& Schechter, P. L., 1996, ApJ, 470, 172.
\bibitem{} Turok, N., 1989, Phys. Rev. Lett., 51, 1716.
\bibitem{} Vilenkin, A., 1981, Phys. Rev., D23, 852.
\bibitem{} Zanchin, V., Lima, J. A. S., \& Brandenberger, R., 1996, Phys. Rev., D54, 7219; astro-ph/9607062.
\bibitem{} Zel'dovich, Ya. B., 1970, Astron. Astrophys. 5, 84.
\bibitem{} Zel'dovich, Ya. B., 1980, MNRAS, 192, 663.
\end{thebibliography}
\end{document}